\documentclass{emulateapj}
\usepackage{graphicx}
\usepackage[space]{grffile}
\usepackage{latexsym}
\usepackage{textcomp}
\usepackage{longtable}
\usepackage{multirow,booktabs}
\usepackage{amsfonts,amsmath,amssymb}
\usepackage{natbib}
\usepackage{url}
\usepackage{CJKutf8}
\usepackage{hyperref}
\usepackage{enumerate}
\hypersetup{colorlinks=false,pdfborder={0 0 0}}
\newif\iflatexml\latexmlfalse
\usepackage[utf8]{inputenc}
\usepackage[english]{babel}
\usepackage{graphicx, color}

\newcommand\T{\rule{0pt}{2.6ex}}       

\newcommand{\etal}{{et al.}}
\newcommand{\ie}{{\it i.e.}}

\newcommand{\be}{\begin{equation}}
\newcommand{\ee}{\end{equation}}

\shorttitle{Data Release 14}
\shortauthors{SDSS Collaboration}

\begin{document}

\title{The Fourteenth Data Release of the Sloan Digital Sky Survey: First Spectroscopic Data from the Extended Baryon Oscillation Spectroscopic Survey and from the Second phase of the Apache Point Observatory Galactic Evolution Experiment}

\email{spokesperson@sdss.org}
\author{Bela Abolfathi\altaffilmark{1},
D. S. Aguado\altaffilmark{2,3},
Gabriela Aguilar\altaffilmark{4},
Carlos Allende Prieto\altaffilmark{2,3},
Andres Almeida\altaffilmark{5},
Tonima Tasnim Ananna\altaffilmark{6},
Friedrich Anders\altaffilmark{7},
Scott F.~Anderson\altaffilmark{8},
Brett H.~Andrews\altaffilmark{9},
Borja Anguiano\altaffilmark{10}
% ...
}
\altaffiltext{1}{Department of Physics and Astronomy, University of California, Irvine, Irvine, CA 92697, USA}
\altaffiltext{2}{Instituto de Astrof\'isica de Canarias, E-38205 La Laguna, Tenerife, Spain}
\altaffiltext{3}{Departamento de Astrof\'isica, Universidad de La Laguna (ULL), E-38206 La Laguna, Tenerife, Spain}
\altaffiltext{4}{Instituto de Astronom{\'i}a, Universidad Nacional Aut\'onoma de M\'exico, A.P. 70-264, 04510, M\'exico, D.F., M\'exico}
\altaffiltext{5}{Instituto de Investigaci\`on Multidisciplinario en Ciencia y Technolog\`ia, Universidad de La Serena, Benavente 980, La Serena, Chile}
\altaffiltext{6}{Yale Center for Astronomy and Astrophysics, Yale University, New Haven, CT, 06520, USA}
\altaffiltext{7}{Leibniz-Institut f\"ur Astrophysik Potsdam (AIP), An der Sternwarte 16, D-14482 Potsdam, Germany}
\altaffiltext{8}{Department of Astronomy, Box 351580, University of Washington, Seattle, WA 98195, USA}
\altaffiltext{9}{PITT PACC, Department of Physics and Astronomy, University of Pittsburgh, Pittsburgh, PA 15260, USA}
\altaffiltext{10}{Department of Astronomy, University of Virginia, 530 McCormick Road, Charlottesville, VA 22904-4325, USA}
% ...

\begin{abstract}
The fourth generation of the Sloan Digital Sky Survey (SDSS-IV) has
been in operation since July 2014. This paper describes the second
data release from this phase, and the fourteenth from SDSS overall
(making this, Data Release Fourteen or DR14). This release makes
public data taken by SDSS-IV in its first two years of operation (July
2014--2016). Like all previous SDSS releases, DR14 is cumulative, including the most recent reductions
and calibrations of all data taken by SDSS since the first
phase began operations in 2000. New in DR14 is the first public release of data from the extended Baryon Oscillation Spectroscopic Survey (eBOSS); the first data from the second phase of the Apache Point Observatory (APO) Galactic Evolution Experiment (APOGEE-2), including stellar parameter estimates from an innovative data driven machine learning algorithm known as ``The Cannon"; and almost twice as many data cubes from the Mapping Nearby Galaxies at APO (MaNGA) survey as were in the previous release ($N = 2812$ in total).  This paper describes the location and format of the publicly available data from SDSS-IV surveys. We provide references to the important technical papers describing how these data have been taken (both targeting and observation details) and processed for scientific use. The SDSS website (www.sdss.org) has been updated for this release, and provides links to data downloads, as well as tutorials and examples of data use.  SDSS-IV is planning to continue to collect astronomical data until 2020, and will be followed by SDSS-V.

\end{abstract}


\section{Introduction}
\label{sec:intro}
\setcounter{footnote}{0}

It is now sixteen years since the first data release from the Sloan Digital Sky Survey (SDSS; \citealt{york00a}). This Early Data Release, or EDR, occured in June 2001 (\citealt{2002AJ....123..485S}). Since this time, annual data releases from SDSS have become part of the landscape of
astronomy, making the SDSS's 2.5 meter Sloan Foundation Telescope \citep{Gunn2006} one of the most productive observatories in the world \citep{2009BAAS...41..913M}, and populating databases used by thousands of astronomers worldwide \citep{Raddick2014a,Raddick2014b}. This paper describes the fourteenth public data release
from SDSS, or DR14, released on 31st July 2017.

The SDSS has completed three phases and is currently in its fourth
phase. SDSS-I and -II conducted a Legacy survey of galaxies and
quasars (\citealt{york00a}), the SDSS-II Supernova Survey
(\citealt{frieman08b, sako14a}), and conducted observations of stars
for the Sloan Extension for Galactic Understanding and Exploration 1
(\mbox{SEGUE-1;} \citealt{yanny09a}). These surveys made use of the SDSS
imaging camera (\citealt{gunn98a}) and 640-fiber optical spectrograph
(\citealt{smee13a}). SDSS-III continued observations of stars with
SEGUE-2, and conducted two new surveys with new instrumentation \citep{2011AJ....142...72E}. 

The Baryon Oscillation Spectroscopic Survey (BOSS; \citealt{dawson13a}) upgraded the optical spectrograph to 1000 fibers (named the BOSS spectrograph; \citealt{smee13a}) to conduct a large volume cosmological redshift survey which built on the work of both SDSS-II (York et al. 2000) and 2dFGRS (Colless et al. 2003).  At the same time, the Apache Point Observatory Galactic Evolution Experiment 1 (APOGEE-1; \citealt{majewski15a}) employed a high resolution near-infrared spectrograph to observe stars in the Milky  Way. All of these observations were conducted at Apache Point
Observatory, and data were publicly released in DR12 (\citealt{alam15a}).

This paper contains new data and data
reductions produced by SDSS-IV (\citealt{blanton17a}). SDSS-IV began observations 
in July 2014, and consists of three programs. 

\begin{enumerate}
\item The extended Baryon
Oscillation Spectroscopic Survey (eBOSS; \citealt{dawson15a}) is
surveying galaxies and quasars at redshifts $z\sim 0.6$--$3.5$ for large scale
structure. eBOSS covers a wider class of galaxies than BOSS at higher effective redshifts. In particular the size and depth of the quasar sample is a huge leap forwards over previous surveys. eBOSS will also observe Emission Line Galaxies, extending the WiggleZ survey (Blake et al. 2011) in the Southern Sky to a larger sample of galaxies at higher redshifts. Following on from eBOSS, the TAIPAN survey (da Cunha et al. 2017) will soon provide a low-redshift complement in the Southern hemisphere. All of these surveys will be eclipsed by forthcoming experiments including DESI (Aghamousa  et al. 2016a,b), Euclid (Laureijs et al 2011) and 4MOST (de Jong et al. 2014), which will use new instrumentation to obtain galaxy surveys an order of magnitude larger than ongoing surveys. Two major subprograms are being conducted concurrently with
eBOSS:  
\begin{itemize}
\item SPectroscopic IDentification of ERosita Sources (SPIDERS)
investigates the nature of $X$-ray emitting sources, including active
galactic nuclei and galaxy clusters. This contains the largest systematic spectroscopic followup sample of X-ray selected clusters (for details see Section 4), reaching into a regime where meaningful dynamical estimates of cluster properties are possible for hundreds of massive systems. It contains a highly complete sample of the most luminous X-ray selected AGN, that will only be superseded by the spectroscopic followup programs of the eROSITA survey (mainly via SDSS-V and 4MOST).
\item Time Domain Spectroscopic Survey
(TDSS; \citealt{morganson15a}) is exploring the physical nature of time-variable sources through spectroscopy. The main TDSS program of optical followup of variables from Pan-STARRS1 imaging is the first large -- by order(s) of magnitude -- program of optical spectroscopy of photometrically variable objects, selected without {\it a priori} restriction based on specific photometric colors or light-curve character. About $20\%$ of TDSS targets involve repeat spectroscopy of select classes of known objects with earlier epochs of spectroscopy, e.g., searching for variability among known BAL quasars, and build and expand on earlier such programs (e.g., Filiz Ak et al. 2014); a comprehensive description of the latter such TDSS repeat spectroscopy programs may be found in MacLeod et al. (2017).
\end{itemize}
 
\item Mapping Nearby Galaxies at APO (MaNGA; \citealt{bundy15a}) is using integral field spectroscopy (IFS) to study 10,000 nearby galaxies. MaNGA builds on a number of successful IFS surveys (e.g., ATLAS-3D, Capellari et al. 2011; DiskMass, Bersady et al. 2010; and CALIFA, S\'{a}nchez et al. 2012) surveying a significantly larger and more diverse samples of galaxies over a broader spectral range at higher spectral resolution. It has finer spatial sampling and a final sample size three times that of the similar SAMI survey (Bryant et al. 2015), and in this release becomes the largest set of public IFS observations available.

\item APOGEE/APOGEE-2 perform a large-scale and systematic investigation of the entire Milky Way Galaxy with near-infrared, high-resolution, and multiplexed instrumentation. For APOGEE-2, observations are being carried out at both Northern and Southern Hemisphere locations: the 2.5m Sloan Foundation Telescope of the Apache Point Observatory (APOGEE-2N; which started Q3 2014) and the 2.5m du Pont Telescope of the Las Campanas Observatory (APOGEE-2S; from Q2 2017). APOGEE/APOGEE-2 is the only large scale ($>$1,000,000 spectra for $>$450,000 objects) NIR spectroscopic survey of stars,  ensuring it has a unique view of all parts of our Galaxy, unhampered by interstellar obscuration in the Galactic plane. Most stellar surveys of equivalent scale -- including those which have concluded (e.g., RAVE, SEGUE-1 and -2, and ARGOS; Steinmetz et al. 2006, Yanny et al. 2009, Rockosi et al. 2009, Freeman et al. 2013), are currently underway (e.g., LAMOST, {\it Gaia}-ESO, GALAH, and {\it Gaia}; Cui+2012, Gilmore et al. 2012, Zucker et al. 2012, Perryman et al. 2001), or are anticipated in the future (e.g., WEAVE, 4MOST, and MOONS; Dalton et al. 2014, de Jong et al. 2014, Cirasuolo et al. 2014) -- have been or will be performed in the optical, and/or with largely medium spectral resolution (however we note plans for a high-res modes for some of these). All of these projects provide complementary data in the form of different wavelength or spatial regimes providing essential contributions to the ongoing census of the Milky Way's stars.

\end{enumerate}


SDSS-IV has had one previous data release (DR13;
\citealt{albareti17a}; for a ``behind the scenes" view of how this is done see \citealt{2016arXiv161205668W}), which contained the first year of MaNGA data,
new calibrations of the SDSS imaging data set, and new processing of APOGEE-1 and BOSS data (along with a small amount of BOSS-related data taken during SDSS-IV). 

DR14 contains new reductions and new data for all programs, roughly
covering the first two years of SDSS-IV operations. This release contains the first public release of data from eBOSS and APOGEE-2, and almost doubles the number of data cubes publicly available from MaNGA.

The full scope of the data release is described in Section 2, and information on data distribution is given in Section 3. Each of the sub-surveys is described in its own section, with eBOSS (including SPIDERS and TDSS) in Section 4, APOGEE-2 in Section 5, and MaNGA in Section 6. We discuss future plans for SDSS-IV and beyond in Section 7.

\section{Scope of Data Release 14}
\label{sec:scope}

As has been the case for all public SDSS data releases, DR14 is cumulative, and includes re-releases of all previously released data processed through the most current data reduction pipelines. In some cases this pipeline has not changed for many DR (see summary below). New data released in DR14 were taken by the Sloan Foundation 2.5m telescope between Aug 23 2014 (MJD= 56893)\footnote{this is the date for eBOSS, for APOGEE and MaNGA it was July 2015} and July 10, 2016 (MJD=57580). The full scope of the release is summarized in Table \ref{table:scope}.

\begin{deluxetable*}{lrr}
\tablewidth{4in}
\tablecaption{Reduced spectroscopic data in DR14 \label{table:scope}} 
\tablehead{ 
\colhead{Target Category} & \colhead{\# DR13} & \colhead{\# DR13+14} }
\startdata
\multicolumn{3}{l}{eBOSS} \\ 
\multicolumn{1}{r}{LRG samples} & 32968 & 138777 \\
\multicolumn{1}{r}{ELG Pilot Survey} & 14459 & 35094  \\
\multicolumn{1}{r}{Main QSO Sample}  & 33928 & 188277 \\	
\multicolumn{1}{r}{Variability Selected QSOs} & 22756 & 87270  \\
\multicolumn{1}{r}{Other QSO samples} & 24840 & 43502  \\
\multicolumn{1}{r}{TDSS Targets} & 17927 & 57675  \\
\multicolumn{1}{r}{SPIDERS Targets} &  3133 & 16394  \\
\multicolumn{1}{r}{Standard Stars/White Dwarfs} &  53584 & 63880  \\
\tableline 
\multicolumn{3}{l}{APOGEE-2} \T \\
\multicolumn{1}{r}{All Stars}  & 164562 & 263444 \\
\multicolumn{1}{r}{NMSU 1-meter stars}  & 894 & 1018 \\
\multicolumn{1}{r}{Telluric stars} & 17293 &  27127 \\
\multicolumn{1}{r}{APOGEE-N Commissioning stars}  & 11917 & 12194 \\
\tableline 
\multicolumn{1}{l}{MaNGA Cubes} &  1390 &  2812 \\ 
\multicolumn{3}{l}{MaNGA main galaxy sample: } \\ 
\multicolumn{1}{r}{\tt PRIMARY\_v1\_2} &  600  & 1278 \\ 
\multicolumn{1}{r}{\tt SECONDARY\_v1\_2} &  473  & 947 \\ 
\multicolumn{1}{r}{\tt COLOR-ENHANCED\_v1\_2} & 216  & 447\\  
MaNGA ancillary targets\tablenotemark{1} &  31 & 121 \\
\tablenotetext{1}{Many MaNGA ancillary targets were also observed as
  part of the main galaxy sample, and are counted twice in this table; some ancillary targets are not galaxies.} 
\enddata
\end{deluxetable*}

We discuss the data released by each of the main surveys in detail below, but briefly, DR14 includes: 
\begin{itemize}
\item Data from 496 new eBOSS plates covering $\sim$2480 square degrees observed from September 2014 to May 2016. We also include data from a transitional project between BOSS and eBOSS called the Sloan Extended Quasar, ELG, and LRG Survey (SEQUELS), designed to test target selection algorithms for eBOSS. The complete SEQUELS dataset was previously released in DR13, however DR14 is the first release for eBOSS. The eBOSS data contain mainly Luminous Red Galaxy (LRG) and Quasar spectra, as well as targets from TDSS and SPIDERS. Twenty-three new eBOSS Emission Line Galaxy (ELG) plates are included in DR14 to test final target selection algorithms.  The full ELG survey started collecting spectra in September 2016 and will be part of a future data release. We include in DR14 the first part of the ELG target catalogue (see Table 2) described in \citet{2017arXiv170400338R}. Other eBOSS value added catalogs (VACs) are also released, namely (1) the redshift measurement and spectral classification catalogue using Redmonster \citep{2016AJ....152..205H}, (2) the quasar catalogue Paris et al. (2017b), and (3) a set of composite spectra of quasars binned on spectroscopic parameters \citep{2016ApJ...833..199J}).
\item APOGEE visit-combined spectra as well as pipeline-derived stellar atmospheric parameters and individual elemental abundances for more than 263,000 stars, sampling all major components of the Milky Way. This release includes all APOGEE-1 data from SDSS-III (Aug 2011-Jul 2014) as well as two years of APOGEE-2 data from SDSS-IV (Jul 2014-Jul 2016). APOGEE VACs include (1) an updated version of the APOGEE red-clump catalog (APOGEE-RC; Pinsonneault et al. in prep.), (2) a cross match between APOGEE and the Tycho-Gaia Astrometric Solution (APOGEE-TGAS; Anders et al. in prep), and (3) a compilation of four different methods to estimate distances to APOGEE stars \citep{2014Schultheis,2016Santiago,2016Wang,2017Q,2017Holtzman}.
\item Data from 166 MaNGA plates, which results in 2812 reconstructed 3D data cubes (for 2744 unique galaxies, primarily from the main MaNGA target sample, but these data also include ancillary targets and $\sim$50 repeat observations). Internally this set of galaxies have been referred to as MaNGA Product Launch-5 (MPL-5); however the reduction pipeline is a different version from that internal release. The new data relative to what was released in DR13 were taken between 13 Aug 2015 (MJD= 57248) and July 10 2016. The MaNGA release also includes two VACs, which provide spatially resolved stellar population and ionized gas properties from {\sc Pipe3D}  (\citealt{2016RMxAA..52...21S,2016RMxAA..52..171S}; see Section \ref{sec:mangavac1}) and {\sc FIREFLY} (\citealt{2017MNRAS.466.4731G}; see Section \ref{sec:mangavac2}). 
\item The largest ever number of SDSS Value Added Catalogues (VACs) produced by scientists in the collaboration -- twelve in total. See Table \ref{table:vac}. 
\item A re-release of the most current reduction of all data from previous versions of SDSS. In some cases the data reduction pipeline hasn't changed for many DRs, and so has not been re-run. The most recent imaging was released in DR13 \citep{albareti17a}; however only the photometric calibrations changed in that release; the astrometry is the same as in DR9 \citep{2012ApJS..203...21A} and the area released and the other aspects of the photometric reduction remain the same as that in DR8 \citep{2011Aihara}. Legacy Spectra (those observed with the SDSS spectrograph) have also not changed since DR8. There have also been no changes to SEGUE-1 or SEGUE-2 since DR9, or MARVELS since DR12 \citep{alam15a}. For DR14 we have re-reduced BOSS spectra using the eBOSS pipeline, where flux calibration has been improved by adding new atmospheric distortion corrections at the per-exposure level \citep{2016ApJ...831..157M} and by employing an unbiased coaddition algorithm.
\end{itemize}

\begin{deluxetable*}{ll}
\tablecaption{Value Added Catalogues New to DR14\label{table:vac}}
\tablehead{\colhead{Description} & \colhead{Reference(s)}}
\startdata
APOGEE:\\
DR14 APOGEE red-clump catalog & \citet{2014ApJ...790..127B} \\
DR14 APOGEE-TGAS Catalogue & Anders \etal ~ in prep.  \\
APOGEE DR14 Distance Estimations from Four Groups & \\
\hspace{2cm} BPG (Bayesian Method) & \citet{2016Santiago,2017Q} \\
\hspace{2cm} NAOC (Bayesian Method) & \citet{2016Wang}\\
\hspace{2cm} NICE (Isochrone Matching Technique) & \citet{2014Schultheis}\\
\vspace{0.1cm} \hspace{2cm} NMSU (Bayesian Method) & \citet{2017Holtzman}  \\
\hline 
eBOSS/TDSS/SPIDERS:\\
Redshift Measurement and Spectral Classification Catalog with Redmonster & \citet{2016AJ....152..205H}  \\
eBOSS: Emission Line Galaxy (ELG) Target Catalog & \citet{2017arXiv170400338R} \\
FIREFLY Stellar Population Models of SDSSI-III and eBOSS galaxy spectra &   \citet{comparat2017} \\
The SDSS DR14 Quasar Catalog & Paris et al. 2017b.  \\
Composite Spectra of BOSS Quasars Binned on Spectroscopic Parameters from DR12Q & \citet{2016ApJ...833..199J} \\
SPIDERS X-ray galaxy cluster catalogue for DR14 & \citet{clerc_etal:16} \\
The Brightest Cluster Galaxy properties of SPIDERS X-ray galaxy clusters & Erfanianfar et al. in prep. \\
Multiwavelength properties of RASS AGN & Merloni et al., in prep \\
Multiwavelength properties of XMMSL AGN & Del Moro et al., in prep. \\
\hline
MaNGA:\\
MaNGA Pipe3D: Spatially resolved and integrated properties of galaxies &  \citet{2016RMxAA..52...21S,2016RMxAA..52..171S,2017arXiv170905438S} \\
MaNGA FIREFLY Stellar Populations & \citet{2017MNRAS.466.4731G}  
\enddata
\end{deluxetable*}

\section{Data Distribution}

The DR14 data are distributed through the same mechanisms as DR13, with the addition of a web application to interactively interface with optical and infrared spectra. We describe our three distribution mechanisms below. These methods are also documented on the SDSS website (\url{http://www.sdss.org/dr14/data\_access}), and tutorial examples for accessing and working with SDSS data can be found at \url{http://www.sdss.org/dr14/tutorials}.

The raw and processed imaging and spectroscopic data, as well as the value added catalogs, are available through the Science Archive Server (SAS, \url{data.sdss.org/sas/dr14}). Data can be downloaded from the SAS directly by browsing the directory structure, and also in bulk using rsync, wget and Globus Online (see \url{http://www.sdss.org/dr14/data_access/bulk} for more details). The data files available on the SAS all have their own datamodel, which describes the content of each file in detail. These datamodels are available at \url{https://data.sdss.org/datamodel}.

The processed imaging and optical and infrared spectra on the SAS are
also available through an interactive web application
(\url{https://dr14.sdss.org}). This web application allows the user to
search for spectra based on specific paameters, e.g. plate, redshift,
coordinates, or observing program. Searches can be saved through
permalinks and options are provided to download the spectra directly
from the SAS, either individually or in bulk. Previous data releases
back to DR8 are available through the same interface. A link is also
provided to the SkyServer explore page for each object.

Finally, the DR14 data can be found on the Catalog Archive Server (CAS, \citealt{2008CSE....10...30T,2008CSE....10....9T}). The CAS stores catalogs of photometric, spectroscopic and derived quantities; these are available through the SkyServer web application (\url{http://skyserver.sdss.org}) for browser-based queries in synchronous mode, and through CasJobs (\url{http://skyserver.sdss.org/casjobs}), which offers more advanced and extensive query options in asynchronous or batch mode, with more time-consuming queries able to run in the background \citep{2008CSE....10...18L}. The CAS is part of the SciServer (\url{http://www.sciserver.org}) collaborative science framework, which provides users access to a collection of data-driven collaborative science services, including SkyServer and CasJobs. Other services include SciDrive, a ``drag-and-drop" file hosting system that allows users to share files; SkyQuery, a database system for cross-matching astronomical multi-wavelength catalogs; and SciServer Compute, a system that allows users to upload analysis scripts as Jupyter notebooks (supporting Python, MatLab and R) and run these databases in Docker containers.

In addition to the data, the data processing software used by the APOGEE-2, eBOSS, and MaNGA teams to derive their data products from the raw frames, is available at \url{http://www.sdss.org/dr14/software/products}.

\section{eBOSS, TDSS and SPIDERS}
\label{sec:eboss}

The extended Baryon Oscillation Spectroscopic Survey (eBOSS;
\citealt{dawson15a}) is surveying galaxies and quasars at redshifts $z
\sim 0.6-3.5$ to map the large scale structure of the Universe with
the main goal to provide Baryonic Acoustic Oscillation (BAO)
measurements in the uncharted redshift change spanning $0.6<z<2.2$. eBOSS achieves this by observing a new set of targets: high redshift LRGs, ELGs, and quasars. The
three new tracers will provide BAO distance measurements with a
precision of 1\% at $z=0.7$ (LRGs), 2\% at $z=0.85$ (ELGs), and 2\% at
$z = 1.5$ (quasars). The Lyman-$\alpha$ forest imprinted on
approximately 120,000 new quasar spectra will give eBOSS an improved
BAO measurement of 1.4$\times$ over that achieved by BOSS
\citep{2015A&A...574A..59D,bautista2017}. Furthermore, the clustering
from eBOSS tracers will allow new measurements of redshift-space
distortions (RSD), non-Gaussianity in the primordial density field,
and the summed mass of neutrino species. eBOSS will provide the first
percent-level distance measurements with BAO in the redshift range
$0.6 < z < 3$, when cosmic expansion transitioned from deceleration to
acceleration. The new redshift coverage of eBOSS obtained by targeting
three classes of targets (LRG, ELG and Quasars) will have the statistical
power to improve constraints relative to BOSS by up to a factor of 1.5
in $\Omega_M$, a factor of three in the Dark Energy Task Force Figure of
Merit (\citealt{albrecht06a}), and a factor of 1.8 in the sum of the
neutrino masses (\citealt{zhao16a}).

\begin{figure}
\centering
\includegraphics[angle=0,width=9.5cm]{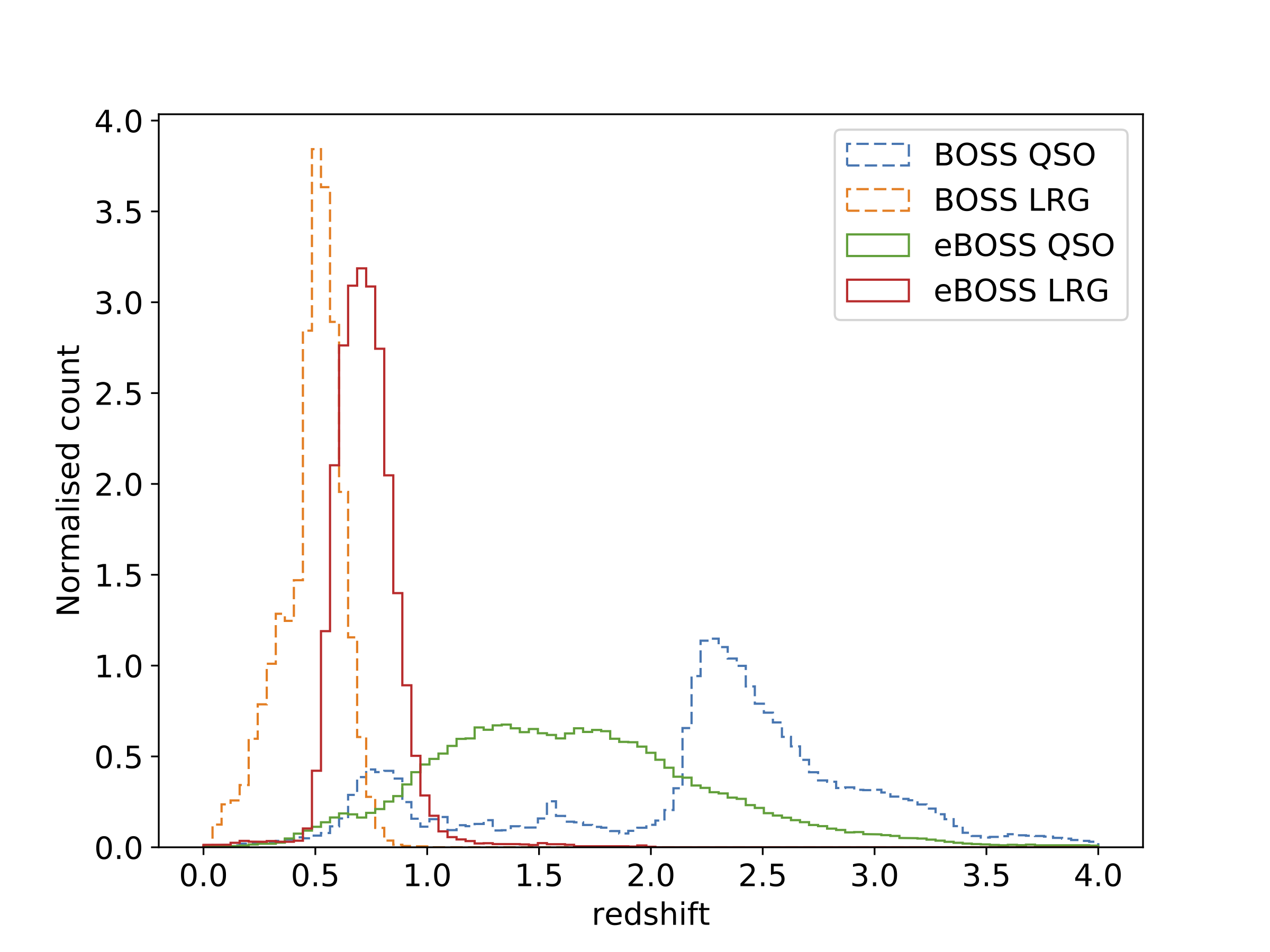}
\caption{$N(z)$ of eBOSS DR14 QSO and LRG compared to DR12 BOSS, demonstrating how eBOSS is filling in the redshift desert between $z\sim 0.8-2.0$. Note that this shows only QSO and LRG targets because no significant number of ELGs have been released in DR14. To convert from Normalized Count to number as a function of redshift, multiply by the total numbers (given in Table 1 for the full survey) times the bin size of $dz= 0.04$.}
\label{fig:ebossnz}
\end{figure}

We show in Figure \ref{fig:ebossnz} the $N(z)$ in eBOSS DR14 QSO and LRG targets compared to the final BOSS release in DR12 \citep{alam15a}, demonstrating how eBOSS is filling in the redshift desert between $z\sim 0.8-2.0$. DR14 does not contain any significant number of ELG targets, which will be released in future DRs.

A significant number of fibers on the eBOSS plates are devoted to two additional dark-time programs. TDSS (Time Domain Spectroscopic Survey; \citealt{morganson15a}) seeks to understand the nature of celestial variables by targeting objects that vary in combined SDSS DR9 and Pan-STARRS1 data (PS1; \citealt{2002SPIE.4836..154K}). A large number of the likely TDSS quasar targets are also targeted by the main eBOSS algorithms and therefore meet the goals of both surveys. TDSS-only targets fill $\sim$10 spectra per square degree.  SPIDERS (Spectroscopic Identification of eROSITA Sources) aims to characterize a subset of X-ray sources identified by eROSITA (extended Roentgen Survey with an Imaging Telescope Array; \citealt{2014SPIE.9144E..1TP}). However, until the first catalog of eROSITA sources is available, SPIDERS will target sources from the RASS (Roentgen All Sky Survey; \citealt{1999A&A...349..389V}) and XMM-Newton (X-ray Multi-mirror Mission; \citealt{2001A&A...365L...1J}). SPIDERS will also obtain on average $\sim$10 spectra per square degree over the course of SDSS-IV, but the number of fibers per square degree on a plate is weighted toward the later years to take advantage of the new data from eROSITA. 

A small fraction of eBOSS time is dedicated to an ancillary program to perform multi-object reverberation mapping for a single 7 deg$^2$ field. This program (SDSS-RM) aims to detect the lags between the broad-line flux and continuum flux in quasars over a broad range of redshift and luminosity with spectroscopic monitoring, which allows the measurement of the masses of these quasar black holes. Started as an ancillary program in SDSS-III, SDSS-RM continues in SDSS-IV with $\sim 12$ epochs (each at nominal eBOSS depth) per year to extend the time baseline of the monitoring and to detect lags on multi-year timescales. The details of the SDSS-RM program can be found in \citet{2015ApJS..216....4S}, and initial results on lag detections are reported in \citet{Shen2016} and \citet{Grier2017}.

eBOSS started in September 2014 by taking spectra of LRG and Quasars, while further development on the definition of the ELG targets sample was conducted in parallel. In May 2016, eBOSS completed its first major cosmological sample containing LRG and Quasars from the first two years of eBOSS data and from SEQUELS (already part of the DR13 release). These data have already been used to improve the classification of galaxy spectra \citep{2016AJ....152..205H}, introduce new techniques to the modeling of incompleteness in galaxy clustering, and to provide measurements of clustering on BAO scales at $1 < z < 2$ for the first time \citep{Ata17}. 

\subsection{Data description}

DR14 includes the data from 496 plates observed under the eBOSS program; it also includes the 126 SEQUELS plates (already released in DR13), from an ancillary program to take advantage of some of the dark time released when BOSS was completed early. The SEQUELS targets are similar to the eBOSS targets as it was a program to test the selection algorithms of eBOSS, in particular the LRG \citep{prakash_etal:16} and quasar algorithms \citep{myers_etal:15}. The final ELG target recipe is not following the one tested during SEQUELS. The new ELG recipe is documented in the DR14 release following the description given by \citet{2017arXiv170400338R}.

For the TDSS program, combined SDSS DR9 and Pan-STARRS1 data (PS1; \citealt{2002SPIE.4836..154K}) are used to select variable object targets \citep{morganson15a,2016ApJ...826..188R}; while for SPIDERS, the objects are selected from a combination of X-ray and optical imaging for the SPIDERS cluster \citep{clerc_etal:16} and AGN \citep{dwelly_etal:17} programs.

\begin{figure*}
\centering
\includegraphics[angle=0,width=15cm]{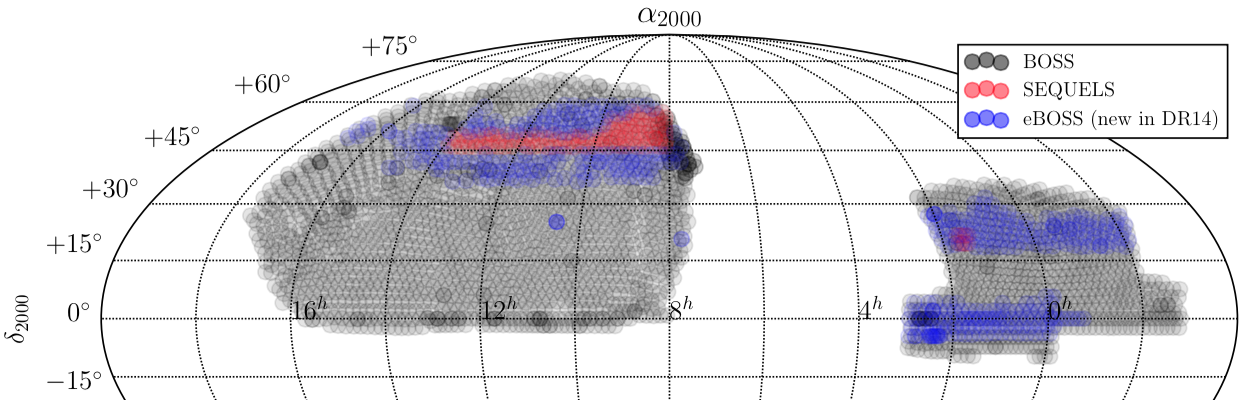}
\caption{DR14 eBOSS spectroscopic coverage in Equatorial coordinates (map centered at RA = 8h.) BOSS coverage is shown in grey, SEQUELS in red, and the eBOSS data newly released for DR14 is shown in blue.}
\label{fig:ebosssky}
\end{figure*}

The sky distribution of the DR14 data from eBOSS is shown in Figure \ref{fig:ebosssky}. Table \ref{table:eboss} summarizes the content and gives brief explanations of the targeting categories. 

\begin{deluxetable*}{llrl}
\tablecaption{eBOSS spectroscopic target categories in DR14 \label{table:eboss}} 
\tablehead{ 
\colhead{Target Category} & \colhead{Target Flag} & \colhead{\# DR14} & \colhead{Reference(s)}}
\startdata
Main LRG sample & {\tt                 LRG1\_WISE   } & 105764 & \cite{prakash_etal:16} \\
Ancillary LRG sample & {\tt                LRG1\_IDROP   } &              45 & \cite{prakash_etal:16}   \\
Main QSO selection & {\tt           QSO1\_EBOSS\_CORE   } &          154349 & \cite{myers_etal:15}\\
Variability-selected QSOs & {\tt              QSO1\_VAR\_S82   } &           10477 & \cite{palanque_etal:16}\\
 & {\tt                  QSO1\_PTF   } &           54037 & \cite{myers_etal:15} \\
Reobserved BOSS QSOs & {\tt                QSO1\_REOBS   } &           16333 & \cite{myers_etal:15} \\
& {\tt             QSO1\_BAD\_BOSS   } &             584 & \\
QSOs from FIRST survey & {\tt          QSO1\_EBOSS\_FIRST   } &            1792 & \cite{myers_etal:15}\\
All eBOSS QSOs also in BOSS & {\tt           QSO\_BOSS\_TARGET   } &             583 & \cite{myers_etal:15}\\
All eBOSS QSOs also in SDSS & {\tt           QSO\_SDSS\_TARGET   } &              20 & \cite{myers_etal:15} \\
All ``known" QSOs & {\tt                 QSO\_KNOWN   } &              11 & \cite{myers_etal:15} \\
Time-domain spectroscopic survey (TDSS) & {\tt               TDSS\_TARGET   } &           39748 & \cite{morganson15a, MacLeod_etal_2017}\\
X-ray sources from RASS \& XMM-Newton & {\tt            SPIDERS\_TARGET   } &           13261 & \cite{clerc_etal:16, dwelly_etal:17} \\
X-ray sources in Stripe 82 & {\tt                S82X\_TILE1   } &            2775 & LaMassa et al 2017 (also see \citealt{lamassa_etal:16}) \\
& {\tt                S82X\_TILE2   } &            2621 & \\
& {\tt                S82X\_TILE3   } &               4  & \\
ELG Pilot Survey & {\tt                 ELG\_TEST1   } &           15235 & \cite{2017MNRAS.465.1831D, Raichoor2016} \\
& {\tt                ELG1\_EBOSS   } &            4741  \\
& {\tt             ELG1\_EXTENDED   } &             659  \\
Standard stars & {\tt                 STD\_FSTAR   } &            8420 & \cite{dawson15a} \\
Standard white dwarfs & {\tt                    STD\_WD   } &             546 & \cite{dawson15a} \\
\end{deluxetable*}

\subsection{Retrieving eBOSS data}

All SDSS data releases are cumulative and therefore the eBOSS data also includes the SEQUELS data taken in SDSS-III or SDSS-IV, reduced with the latest pipelines. eBOSS targets can be identified using the EBOSS\_TARGET1 bitmask. The summary {\tt spAll-v5\_10\_0.fits} datafile, which includes classification information from the pipeline, is located on the SAS\footnote{\tt https://data.sdss.org/sas/dr14/eboss/spectro/redux/v5\_10\_0/}; the data can also be queried via the {\tt specObjAll} table on the CAS.

\subsection{eBOSS/TDSS/SPIDERS VACs} 

We include seven VACs based on BOSS, eBOSS, TDSS or SPIDERS data or target selection in this DR. Brief details of each are given below, and for more details we refer you to the relevant papers in Table \ref{table:vac}. 

\subsubsection{Redshift Measurement and Spectral Classification Catalog with Redmonster} 

The redmonster software\footnote{\tt https://github.com/timahutchinson/redmonster} is a sophisticated and flexible set of Python utilities for redshift measurement, physical parameter measurement, and classification of one-dimensional astronomical spectra. A full description
of the software is given in \citet{2016AJ....152..205H}. The software approaches redshift measurement and classification as a $\chi^2$ minimization problem by cross-correlating the observed spectrum with a theoretically-motivated template within a spectral template class over a discretely sampled redshift interval. In this VAC the software has been run on all DR14 LRG spectra. Redmonster was able to successfully measure redshifts for $\sim$90\% of LRG spectra in DR14. This is an increase of $\sim$15\%, in absolute terms, over {\tt spectro1d}, and nearly matches the most optimistic estimate for the fraction of measurable redshifts as determined by visual inspections. 

\subsubsection{The SDSS-IV Extended Baryon Oscillation Spectroscopic Survey: Emission Line Galaxy Target Catalog}   
 We publish the south galactic cap Emission Line Galaxy (ELG) catalog used for eBOSS \citep{2017arXiv170400338R}. Targets were selected using photometric data from the Dark Energy Camera Legacy Survey (DECaLS; {\tt http://legacysurvey.org/}). We selected roughly 240 ELG targets per square degree. The great majority of these ELG lie in the redshift range $0.67 < z < 1.1$ (median redshift 0.85). 

\subsection{FIREFLY Stellar Population Models of SDSSI-III and eBOSS galaxy spectra}

We determine the stellar population properties - age, metallicity, dust reddening, stellar mass and the star formation history - for all spectra classified as galaxies that were published in this release (including those from SDSSI-III).
We perform full spectral fitting on individual spectra, making use of high spectral resolution stellar population models published in \citet{2017MNRAS.472.4297W}.
Calculations are carried out for several choices of the model input, including three stellar initial mass functions and three input stellar libraries to the models.
We study the accuracy of parameter derivation, in particular the stellar mass, as a function of the signal-to-noise of the galaxy spectra. We find that signal to noise ratio per pixel around 20 (5) allow a statistical accuracy on $\log_{10}(M^{*}/M_{\odot})$ of 0.2 (0.4) dex, for the Chabrier IMF. We publish all catalogs of properties as well as model spectra of the continuum for these galaxies\footnote{\url{https://www.sdss.org/dr14/spectro/eboss-firefly-value-added-catalog}} \citep{comparat2017}. This catalog is about twice as large as its predecessors (DR12) and will be useful for a variety of studies on galaxy evolution and cosmology.

\subsubsection{The SDSS DR14 Quasar Catalog}

Following the tradition established by SDSS-I/II/III the SDSS-IV/eBOSS collaboration is producing a visually inspected quasar catalog. The SDSS-DR14 quasar catalog (DR14Q; Paris et al. 2017b) is the first to be released that contains new identifications that are mostly from eBOSS. The contents of this are similar to the DR12 version (which contained final data from BOSS as well as data from the preliminary eBOSS survey "SEQUELS") as described in Paris et al. (2017a). 

\subsubsection{Composite Spectra of BOSS Quasars Binned on Spectroscopic Parameters from DR12Q}

We present high signal-to-noise composite spectra of quasars over the redshift range $2.1 \le z \le 3.5$. These spectra, based on the DR12 BOSS quasar catalog \citep{alam15a} are  binned by luminosity, spectral index, and redshift. As discussed in \citet{2016ApJ...833..199J}, these composite spectra can be used to reveal spectral evolution while holding luminosity and spectral index constant. These composite spectra allow investigations into quasar diversity, and can be used to improve the templates used in redshift classification. See Jensen et al. (2017) for more details. 

\subsubsection{SPIDERS X-ray galaxy cluster catalogue for DR14\label{SPIDERSclusters}}

A substantial fraction of SPIDERS fibers target red-sequence galaxies in candidate X-ray galaxy clusters. The systems were found by filtering X-ray photon overdensities in the ROSAT All-Sky Survey (RASS) with an optical cluster finder \citep[see][for details on the samples and targeting strategy]{clerc_etal:16}. Adding together the DR14 eBOSS sky area with the SEQUELS area (Fig. 1), 573 of these systems show a richness $\lambda_{\rm OPT} > 30$, have been completely observed as part of DR14. A complete observation means that all tiled galaxies in a cluster red sequence have got a spectrum in DR14; these clusters must also contain at least one redshift from SDSS-I to -IV in their red-sequence. Systems located at a border of the DR14 footprint, but in the interior of the full eBOSS footprint, will be fully covered through later observations by overlapping plates.

A total of 9,029 valid redshifts were associated with these candidate rich galaxy clusters, leading to a median number of 15 redshifts per red sequence. An automated algorithm performed a preliminary membership assignment and interloper removal based on standard iterative $\sigma$-clipping method. The results of the algorithm were visually inspected by 8 experienced galaxy cluster observers, ensuring at least two independent evaluators per system. A web-based interface was specifically developed to this purpose: using as a starting point the result of the automated algorithm, the tool allows each inspector to interactively assess membership based on high-level diagnostics and figures \citep[see Fig.~16 in][]{clerc_etal:16}. A final decision is made by each evaluator whether to validate the system as a bona-fide galaxy cluster, or ``unvalidate" the system by lack of data or identification of a false candidate. Validation is in most cases a consequence of finding three or more red-sequence galaxies in a narrow redshift window, compatible with them all being galaxy cluster members. A robust weighted average of the cluster members redshifts provides the cluster systemic redshift. A majority vote was required for each system to be finally "validated" or ``unvalidated" ; in the former case, an additional condition for agreement is the overlap of the cluster redshifts 95\% confidence intervals. A second round of evaluations involving four inspectors per system was necessary to resolve cases with no clear majority.

In total, 520 of these systems are validated as true galaxy clusters based on spectroscopic data and they form the SPIDERS X-ray galaxy cluster Value-Added catalogue for DR14. Among them 478 are unique components along a line-of-sight. A total of 7,352 spectroscopic galaxies are members of a galaxy cluster. This catalogue in particular lists each galaxy cluster redshift and its uncertainty, its number of spectroscopic members and its X-ray luminosity, assuming each component along a line-of-sight contributes the flux measured in RASS data.

\subsubsection{The Brightest Cluster Galaxy properties of SPIDERS X-ray galaxy clusters}

We provide the brightest cluster galaxies (BCGs) catalog for the SPIDERS DR14 X-ray detected galaxy clusters VAC (see Section ref{SPIDERSclusters}). BCGs have been identified based on the available spectroscopic data from SPIDERS and photometric data from SDSS (Erfanianfar et al. in prep.). Only those SPIDERS clusters which have one component in the SPIDERS X-ray galaxy clusters are considered in this analysis. Stellar masses and SFRs of the BCGs are computed by combining SDSS, WISE \citep{Lang2014a, Lang2014b, Meisner2017} and GALEX \citep{Budavari2009} photometry, and using state-of-the-art spectral energy distribution (SED) fitting \citep{Ilbert2006,1999MNRAS.310..540A}. Where available, the star formation rate is taken from the MPA-JHU galaxies properties VAC \citep{2004MNRAS.351.1151B} instead of from the SED fitting. The structural properties (effective radius, Sersic index, axis ratio and integrated magnitude) for all BCGs are provided by Sersic profile fitting using SIGMA \citep{2012MNRAS.421.1007K} in three optical bands (g,r,i; Furnell et al. in prep.). This catalog lists the BCGs identified as part of this process, along with their stellar mass, star formation rates (SFR), and
structural properties.

\subsubsection{Multiwavelength properties of RASS and XMMSL AGN }

In  these two VACs, we present the multi-wavelength characterization over the area covered by the SEQUELS and eBOSS DR14 surveys (~2500 deg$^2$) of two highly complete samples of X-ray sources: 
\begin{enumerate}
\item The ROSAT All-Sky Survey (RASS) X-ray source catalogue (2XRS; Boller et al. 2016) 
\item The XMM-Newton Slew Survey point source catalog (XMMSL; Saxton et al. 2008; version 1.6).
\end{enumerate} 
We provide information about X-ray properties of the sources, as well as of their counterparts at longer wavelength (Optical, IR, Radio) identified first in the All-WISE IR catalog\footnote{\tt http://wise2.ipac.caltech.edu/docs/release/allwise/} via a Bayesian cross-matching algorithm (Dwelly et al. 2017; Salvato et al. 2018). We complement this with dedicated visual inspection of the SDSS spectra, providing accurate redshift estimates (with objective confidence levels) and source classification, beyond the standard eBOSS pipeline results.

\section{APOGEE-2}
\label{sec:apogee}

DR14 is the fourth release from the Apache Point Observatory Galactic Evolution Experiment (APOGEE). DR14 presents, for the first time, the first two years of SDSS-IV APOGEE-2 data (Jul 2014-Jul 2016) as well as re-processed data from SDSS-III APOGEE-1 (Aug 2011-Jul 2014). Note that the general term APOGEE data, employed throughout this paper, refers to both APOGEE-1 and APOGEE-2 data.  APOGEE-2 data are substantively the same as APOGEE-1 data, however, one of the three detectors in the instrument was replaced at the end of APOGEE-1 because it exhibited a substantial amount of persistence (i.e. light from previous exposures led to excess recorded charge in subsequent exposures). The new detector is substantially better in this regard.

\begin{figure*}
\centering
\includegraphics[angle=0,width=15cm]{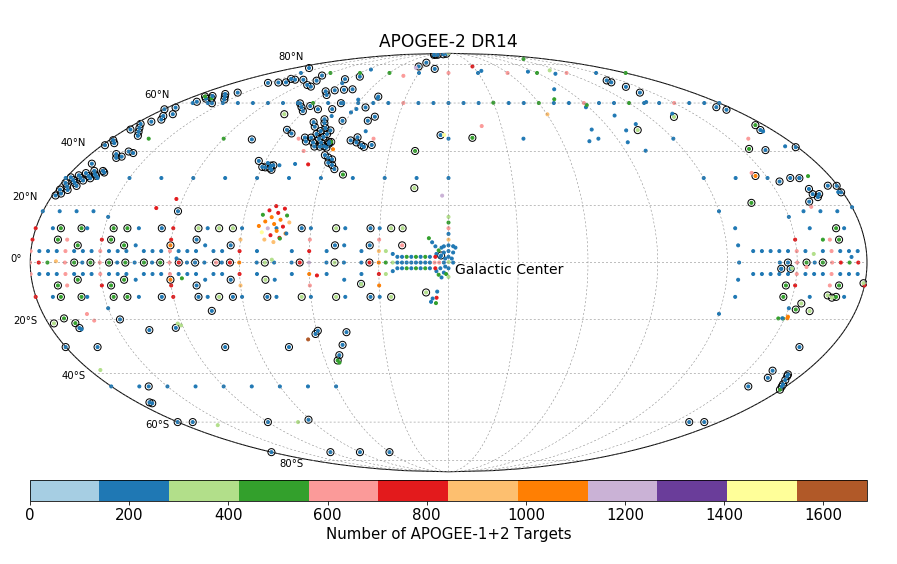}
\caption{DR14 APOGEE spectroscopic coverage in Galactic coordinates (map centered on the Galactic center). The color coding indicates the number of APOGEE-1+2 targets observed per field, as shown in the key. Fields new to DR14 are outlined in black. }
\label{fig:apogeesky}
\end{figure*}

APOGEE data in DR14 includes visit-combined spectra as well as pipeline-derived stellar atmospheric parameters and individual elemental abundances for 263,444 stars\footnote{The figure of 263,444 results from the removal of duplicate observations for a single star.  Note that DR14 has a total of 277,731 entries.}, sampling all major components of the Milky Way.  The DR14 coverage of APOGEE data is shown in Galactic co-ordinates in Figure \ref{fig:apogeesky}.  In addition to the Milky Way bulge, disk, and halo, DR14 includes, for the first-time, data from stars in satellite galaxies, which are typically fainter targets than those from the main portion of the survey.  DR14 incorporates a few modifications in the Data Reduction Pipeline (DRP) as well as in the APOGEE Stellar Parameter and Chemical Abundance Pipeline (ASPCAP).  It also includes a separate set of stellar parameters and abundances from The Cannon (Ness et al. 2015)\footnote{Named in recognition of the stellar classification work of Annie-Jump Cannon \citep{1918AnHar..91....1C}.}.

Two separate papers will provide more in-depth discussion and analysis of APOGEE data released in DR13/DR14: Holtzman et al. (in prep.) describes in detail the DR13/DR14 pipeline processing as well as the associated data products; and J\"onsson et al. (in prep.) compares stellar parameter and element abundances from DR13/DR14 with those from the literature.

\subsection{Targeting}
The targeting strategy of APOGEE-2 departs slightly from that of APOGEE-1 and is set based on three-tier priority scheme: core, goal, and ancillary science \citep{2017arXiv170800155Z}.  The core science targets, which are the highest priority, are those that directly address the primary objectives of APOGEE and include the Galactic bulge, disk, and halo, globular and open clusters, {\it Kepler} field spectroscopic follow-up, and satellite galaxies (unique in APOGEE-2).  ``Goal" science targets fall in line with APOGEE science goals with a second-tier prioritization and include M dwarfs, eclipsing binaries, substellar companions, {\it Kepler} Objects of Interest, young (star-forming) clusters, and Extended {\it Kepler} Mission (K2) spectroscopic follow-up.  The third-tier priority are ancillary science targets, for which a general solicitation was issued for programs that could harness the unique capability of the APOGEE instrument.  

 Since the ancillary programs of APOGEE-1 were largely successful and broadened its scientific scope,  APOGEE-2 continues in this vein and DR14 presents some of the first ancillary program data.  As in APOGEE-1, the primary stellar targets of APOGEE-2 are red giant branch (RGB) stars.  APOGEE-2 extends the target stellar classes with designated observations of red clump (RC) stars in the bulge as well as faint stars (e.g., dwarf Spheroidal and halo stream members with $H \geq 14$).  On top of the APOGEE-led programs, additional data are collected with the MaNGA co-targeting program.  For the MaNGA pointings, APOGEE data is collected concurrently, with the targeted fields in the direction of the Galactic caps.  To document the APOGEE-2 targeting scheme, a new set of bit flags is employed in DR14: \texttt{APOGEE2\_TARGET1, APOGEE2\_TARGET2, and APOGEE2\_TARGET3}.  Further details with regard to the APOGEE-2 targeting strategy and field design may be found in \citet{2017arXiv170800155Z}, including information on APOGEE-2S targets which will are planned to be part of the next data release.

\subsection{Reduction and Analysis Pipeline. Data Products}

\begin{figure}
\centering
\includegraphics[angle=0,width=9.2cm]{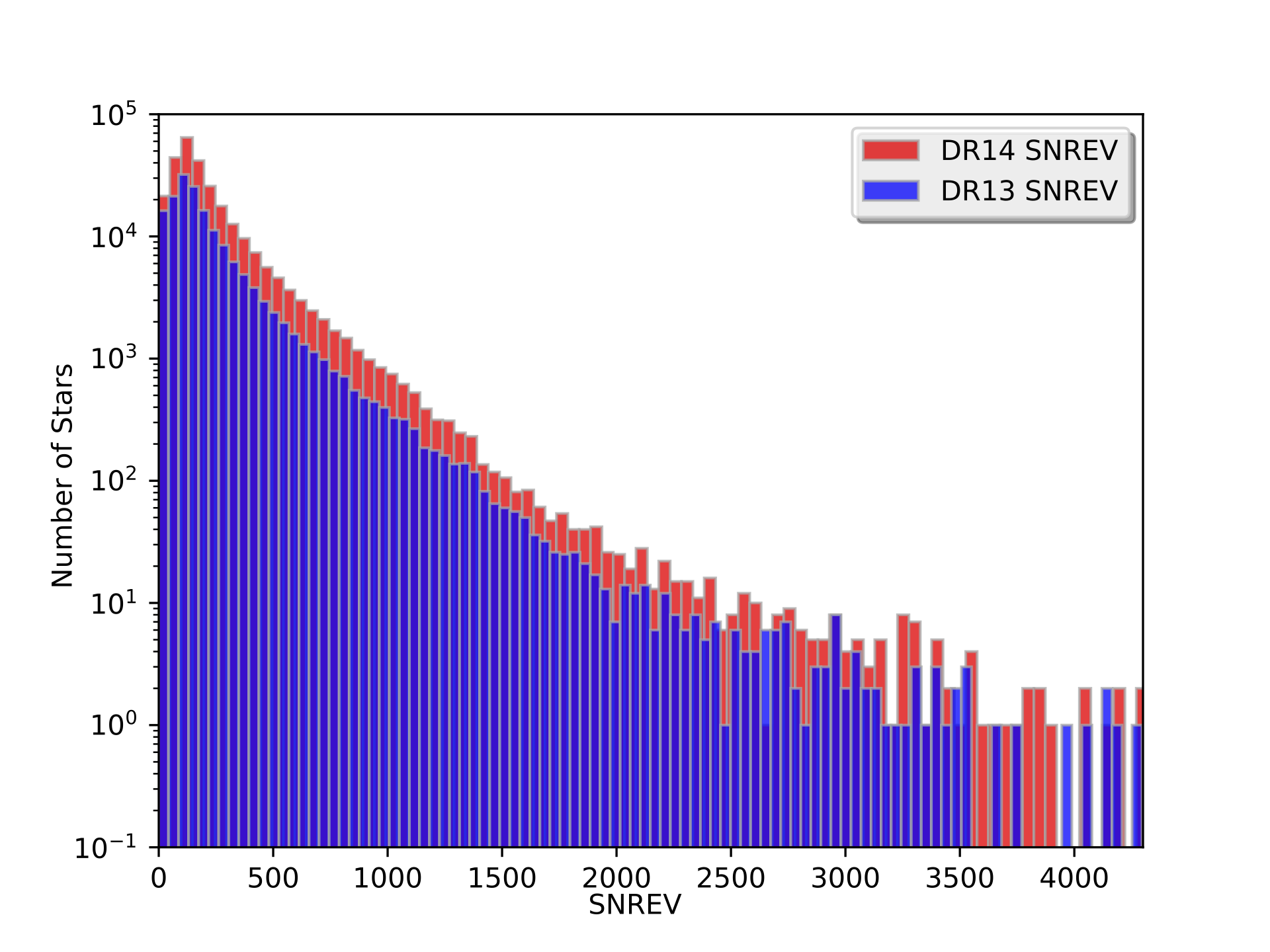}
\caption{A comparison of the $S/N$ distribution of APOGEE spectra released in DR14 (red) with those released in DR13 (blue).  The $S/N$ quantity displayed in the figure is {\tt SNREV}, a revised $S/N$ estimate which considers persistence issues.}
\label{fig:apogeesnr}
\end{figure}

As with the previous data releases, all spectra are processed through the DRP, which includes dark current subtraction, cosmic ray removal, flat-fielding, spectral extraction, telluric correction, wavelength calibration, and dither combination.  Radial velocities (RVs) are determined for each individual visit and the individual visit spectra are resampled to rest-wavelength and combined to generate a single spectrum for each object.  Associated DRP data products are the visit-combined spectra and radial velocity (RV) values.  For DR14, modifications to the RV determination and associated star combination have occurred.  The RV values are now determined both relative to the combined spectrum (in an iterative fashion) as well as to the best-matching model. The radial velocities from the method that yields the lower scatter are adopted (\texttt{VHELIO\_AVG}) and estimates of the associated error and scatter are generated.  Note that the new methodology has resulted in improved RV determinations for low signal-to-noise observations (and consequently, faint stars), but there can still be potentially significant issues with some of the faintest targets. The distribution of $S/N$ values for spectra released in DR14 (compared to those released in DR13; \citealt{albareti17a}) are shown in Figure \ref{fig:apogeesnr}.

\subsubsection{Persistence}

As discussed in \citet{nidever2015} and \citet{holtzman2015}, one of the three APOGEE-1 detectors (the ``blue" detector) exhibited significant levels of persistence (\ie. charge which is held between exposures) over one third of the detector area, and another (the ``green" detector) exhibited persistence at a somewhat lower level over a smaller area. This persistence affected the derived stellar abundances \citep{holtzman2015}. As mentioned above, the ``blue" detector was replaced for APOGEE-2 in part to solve this problem. For the APOGEE-1 data, we attempt to subtract out persistence based on a model and also de-weight pixels affected by persistence during visit combination in such a way that for stars with a mix of persistence-affected and non-persistence-affected visits, the combined spectra are dominated by the non-affected visits. This results in a reduction of the systematic errors, but a slight increase of the random errors. This process significantly reduces the impact of persistence (Holtzman et al. in prep); however, it can still have an effect, especially for fainter targets. Users of the APOGEE spectra should pay careful attention to the pixel-level data flags and the pixel uncertainties.

\subsubsection{ASPCAP}

After the DRP stage, the visit-combined stellar spectra are processed by ASPCAP, which derives the stellar atmospheric parameters (e.g., effective temperature $T{\rm eff}$, surface gravity $\log g$, metallicity $[M/H]$) as well as abundances for more than 20 species.  The ASPCAP determination proceeds in three stages: an initial pass through ASPCAP gives coarse values for a few key atmospheric parameters to identify which spectral grids should be used on each object, a second pass yields the full set of parameters, and then a final pass determines abundances for each element with stellar parameters fixed. 

For DR14, ASPCAP modifications include a new normalization scheme for both observed and synthetic spectra. Rather than using an iteratively asymmetrically-clipped fit, the continuum is determined by a polynomial fit to the spectra after masking of sky lines. This new scheme avoids clipping, since it leads to systematic differences in continuum normalization as a function of S/N. Another change is that the ASPCAP parameter determination was done by $\chi^2$-minimization over a 7-dimensional grid for giants which included a microturbulence dimension.  This leads to slightly lower abundance scatter in clusters as well as smaller trends of $[M/H]$ with temperature.

One caveat of the DR14 ASPCAP analysis is that new grids were not constructed for APOGEE-2 line spread functions (LSFs): grids made with the APOGEE-1 
LSFs were used. Since the only change was the detector replacement, no large LSF changes were expected, nor were they noticed, but subtle differences may be present.

\subsubsection{Calibration and Data Product Usage}

As with previous DRs, DR14 includes a post-ASPCAP calibration of the final stellar atmospheric parameter and element abundances.  A variety of different stellar clusters and standards are employed in the calibration of the results.  These calibrations include a metallicity-dependent temperature correction, a surface gravity calibration based on asteroseismic gravities, an internal and external calibration of metallicity ($[M/H]$), and a temperature-dependent and zero-point calibration for elemental abundances. Note that surface gravity calibration is not done for dwarfs because we do not have independent estimates of surface gravities from which to derive such calibrations. Calibrations are applied to abundances over temperature ranges that are determined by looking at the ranges over which data in star clusters produce the same abundance. Based on cluster results and inspection of the spectra, we do not provide calibrated abundances for Cu, Ge, Y, Rb, and Nd, since these do not appear to be reliable.

Several different bitmasks (\texttt{STARFLAG, PARAMFLAG, ASPCAPFLAG}) are included that provide information on factors that affect data quality, and users are strongly encouraged to pay attention to these   

\subsection{New DR14 Data Product: Results from The Cannon}

New in DR14 is the inclusion of parameters and abundances derived from the Cannon (Ness et al. 2015). The Cannon is a data-driven model that provides parameters and abundances (collectively called labels) from the spectra, after training the sensitivity of each pixel to parameters and abundances based on a training set with independently derived labels. 

For DR14, we train The Cannon on ASPCAP results for a subset of high S/N giant stars, and apply the model to all objects within the range of parameters covered by the training set. DR14 Cannon results have been derived using the Cannon-2 code (Casey et al. 2016), but with a few modifications. First, we adopted uncertainties from the ASPCAP pipeline, which do a better job de-weighting areas around imperfectly subtracted sky lines. 

Second, and more importantly, we use ``censoring" in the derivation of individual elemental abundances, which forces the model to only use pixels where there are known lines of a given element (rather than the full spectrum) to derive the abundance of that element.  This was done because it was discovered that, when using the full spectrum, pixels without known lines of an element (and sometimes, with known lines of another element) contributed to the model sensitivity for that element. This suggests that the model may be affected by correlations of abundances within the training set stars. Without censoring, such correlations can lead to abundances that appear to be of higher precision, but this precision may not reflect higher accuracy, if the correlations are not present over the entire data set. While results for some elements with censoring show less scatter than ASPCAP results, results for other elements can look significantly worse.  The implementation of censoring was done by using the elemental windows used by the ASPCAP analysis; it is possible that this is overly conservative because the ASPCAP windows reject regions in the spectrum that have abundance senstivity if they are also sensistive to other abundances in the same elemental abundance group.

\subsection{APOGEE VACs}

Three APOGEE related VACs are included in DR14. They are briefly summarized below. For more details we refer the reader to the relevent paper in Table \ref{table:vac}.

\subsubsection{DR14 APOGEE red-clump catalog} 

DR14 contains an updated version of the APOGEE red-clump (APOGEE-RC) catalog. This catalog is created using the same procedure as the original APOGEE-RC catalog \citep{2014ApJ...790..127B}  now applied to the ASPCAP parameters derived in this data release. To account for changes in how the ASPCAP-derived $\log g$ is calibrated in DR14, we have made the upper $\log g$ cut more stringent by 0.1 dex (the upper $\log g$ limit in Equation [2] in \citet{2014ApJ...790..127B}  now has 2.4 instead of 2.5). Like in the original release, we also apply an additional $\log g$ cut to remove further contaminants (Equation [9] in \citealt{2014ApJ...790..127B}). Otherwise the catalog is created in the same manner as the original catalog.

The DR14 APOGEE-RC catalog contains 29,502 unique stars, about 50\,\% more than in DR13. Note that because of changes in the target selection in APOGEE-2, the relative number of RC stars in APOGEE-2 is smaller than in APOGEE-1. We provide proper motions by matching to the UCAC-4 \citep{2013AJ....145...44Z} and HSOY \citep{2017A&A...600L...4A} catalogs. Contamination by non-RC stars in the DR14 RC catalog is estimated to be less than 5\,\% by comparing against true RC stars in the APOKASC catalog (M. Pinsonneault et al. in prep.).

\subsubsection{DR14 APOGEE-TGAS Catalogue}

The first data release of the Gaia mission contains improved parallaxes and proper motions for more than 2 million stars contained in the Tycho-2 catalogue, among them 46,033 objects (10,250 of them unique stars) contained in APOGEE DR14. This is known as the  Tycho-Gaia Astrometric Solution (TGAS). We provide the cross-matched catalog, together with precise combined astrometric/spectro-photometric distances and extinctions determined with {\tt StarHorse} \citep{2017Q} for 29,661 stars. We also include orbital parameters calculated using the {\tt GravPot16} code\footnote{https://fernandez-trincado.github.io/GravPot16/} (Fernandez-Trincado et al., in prep.). For more details see Anders et al. (in prep), a summary is also provided in \citet{Anders2017conf}.

\subsubsection{APOGEE DR14 Distance Estimations from Four Groups}

This VAC provides spectro-photometric distance estimates for APOGEE stars that have been calculated by four groups, using slightly different isochrone techniques. All groups used the DR14 calibrated ASPCAP stellar parameters, if they fall inside the calibration ranges (see Holtzman et al., in prep). The distances come from (1) the {\tt StarHorse} code \citep{2016Santiago,2017Q}, (2) the code described in \citet{2016Wang}, (3) the isochrone-matching technique described in \citet{2014Schultheis}, and (4) the distance code described in \citet{2017Holtzman}.

\section{MaNGA}
\label{sec:manga}

In the context of the MaNGA Survey, DR14 roughly doubles the sample size of associated data products that were first made public in DR13.  Spanning observations from the first two years of operations, the DR14 products include raw observations, intermediate reduction output, such as reduced fiber spectra, and final data cubes as constructed by the Data Reduction Pipeline (DRP; \citealt[hereafter L16]{law16}).  A summary {\tt drpall} catalog provides target identification
information, sky positions, and object properties like photometry and redshifts.  The MaNGA observing strategy is described in \cite{law15}, and the flux calibration scheme presented in \cite{yan16}.  An overview of the survey execution strategy and data quality is provided in \citet{yan16a}. \citet{weijmans2016} provides a short summary to the entire survey, which is comprehensively described in \citet{bundy15a}.

\begin{figure*}
\centering
\includegraphics[angle=0,width=15cm]{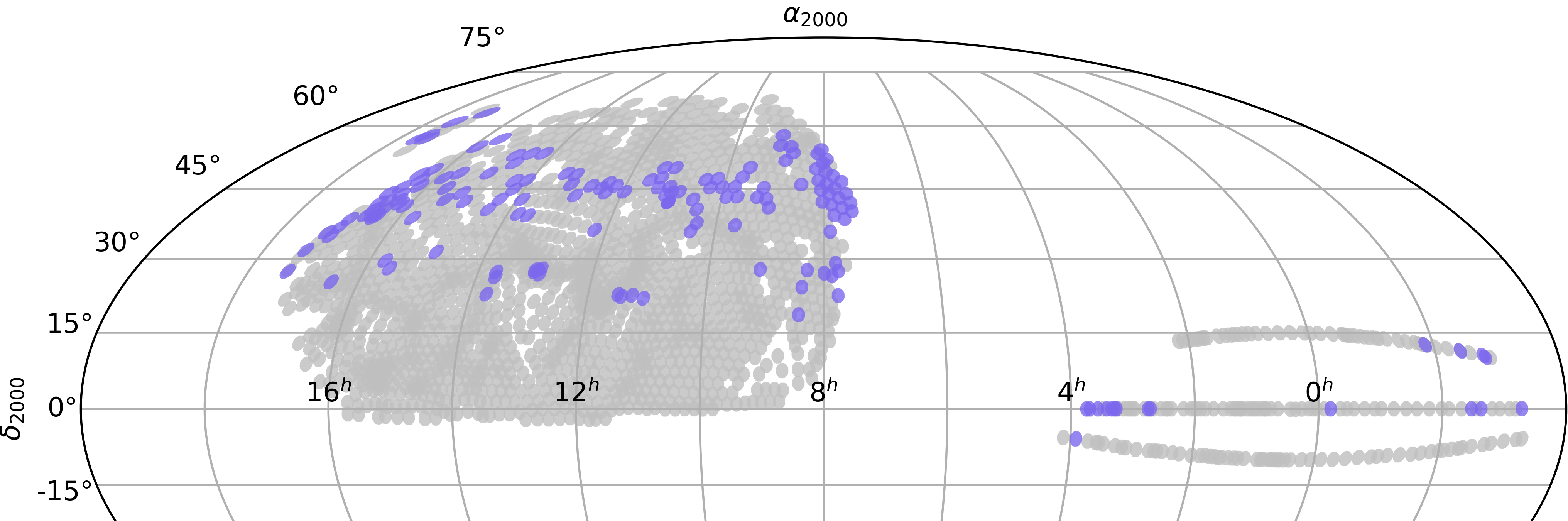}
\caption{The sky distribution (Mollewiede equatorial projection for Dec $>-20^\circ$) of possible MaNGA plates (in grey). Because MaNGA targets are selected from a sample with SDSS-I photometry and redshifts, this footprint corresponds to the Data Release 7 imaging data (Abazajian et al. 2009).  Each plate contains 17 MaNGA targets, and around 30\% of all possible plates will be observed in the full 6-year survey. The purple indicates plates with data released as part of DR14.}
\label{fig:mangasky}
\end{figure*}

DR14 includes observations from 166 MaNGA plates resulting in 2812 datacubes comprising targets in the main samples as well as ancillary programmes, and around 50 repeat observations.  The sky layout of the DR14 released MaNGA data is shown in Figure \ref{fig:mangasky}.

\subsection{MaNGA Target Classes}

The target selection for the MaNGA Survey is described in detail by \citet{Wake2017}.  MaNGA's main galaxy
sample contains galaxies with stellar masses, ${\rm M_*} > 10^9 ~{\rm M}_\odot$, and is comprised of three main sub-samples that are defined on
the basis of SDSS-I/II photometry and spectroscopic redshifts to deliver a final distribution that is roughly flat in
$\log M_{*}$. The Primary sample achieves radial coverage out to 1.5 times the effective radii (1.5 $R_{e}$) for target galaxies, while the Secondary
sample reaches 2.5 $R_e$.  The Color-Enhanced supplement expands the selection of the Primary sample to include
under-represented regions of $M{_*}$-color space.  We refer to the combination of the Primary and Color-Enhanced
supplements as ``Primary+'' which balances the restframe color distribution at fixed $M_{*}$.  The MaNGA samples can be weighted so that they are equivalent to a volume limited sample. The required volume weights are described in \citet{Wake2017} and are provided in the DR14 version of the targeting file.

DR14 includes 1278 Primary galaxies, 947 Secondary galaxies and 447 Color-Enhanced supplement galaxies.  Which sample a
given target galaxy belongs to is given by the MANGA\_TARGET1 bitmask (or mngtarg1 in the ``drpall'' file).  Bits 10,
11, and 12 signal that galaxies were selected as Primary, Secondary, or Color-Enhanced targets respectively.  In
addition to $\sim$121 ancillary program targets, $\sim$50 galaxies were observed as fillers and do not fall into these
target categories.  They should be ignored in statistical studies of the MaNGA data.

MaNGA has also begun observing Milky Way stars in a bright-time survey
program called the MaNGA Stellar Library (MaStar) that makes
use of MaNGA IFUs during APOGEE-2 observations. The goal of MaStar is to build a new stellar library
comprising  $>8000$ stars that span the widest accessible ranges in effective temperature, surface gravity, metallicity, and element abundance ratios
(Yan et al. in prep; see Yan et al 2017 for a conference proceeding describing the plans). Reduced stellar spectra will be included in DR15.

As described in the DR13 paper, roughly 5\% of MaNGA IFUs are allocated to targets defined by approved ancillary programs.  These sources can be identified using the MANGA\_TARGET3 bitmask (or
mngtarg3 in the drpall file).  Most of the programs represented in DR14 are described in DR13\footnote{also see \url{http://www.sdss.org/dr14/manga/manga-target-selection/ancillary-targets}}.  They include targeted followup of AGN hosts, starburst galaxies, merging systems, dwarf galaxies, Milky Way analogs, and brightest cluster galaxies.  New in DR14, we include deep observations reaching $\sim$20 hours in the center of the Coma cluster \citep{2017arXiv170907003G} and IFU observations allocated as part of an ancillary program to a nearby dwarf galaxy that is part of the ACS Nearby Galaxy Survey \citep{2009ApJS..183...67D}. 

\subsection{Working with MaNGA Data}

All MaNGA data products take the form of multi-extension FITS files. As we describe in DR13, the DRP data products consist of intermediate reduced data (sky-subtracted, flux-calibrated fiber spectra with red and blue data
combined for individual exposures of a plate) and final-stage data products (summary row-stacked spectra and data
cubes) for each target galaxy.  The summary row stacked spectra (RSS files) are two-dimensional arrays provided for
each galaxy in which each row corresponds to a single fiber spectrum. 

The three-dimensional data cubes are created by combining the individual spectra for a given galaxy together onto a
regularized 0.5$\arcsec$ grid (see L16 for more detail).  The associated wavelength  arrays for both the data cubes and
RSS files can be accessed in logarithmic and linear scales.  Each data cube contains additional extensions with
information that includes the inverse variance, a bad-pixel mask, instrumental resolution, reconstructed broadband
images, and the effective spatial point-spread function. The full data model for all MaNGA DRP data products can be
found online at \url{http://www.sdss.org/dr14/manga/manga-data/data-model} and in Appendix B of L16.

 The DR14 pipeline for MaNGA is nearly identical to that in DR13 with a few small exceptions listed below:  
\begin{itemize}
\item The spectral resolution reported is worse by about 10\%. This change reflects growing understanding of the data quality to account for the effects of both pre- vs post-pixellization gaussian profile fitting and changes in the line spread function (LSF) introduced by the wavelength rectification. There are likely to be further small change in future data releases. 
\item Local reddening maps (rather than plate averages) have been used in calculations of $S/N$ of the spectra
\item Spaxels flagged as containing foreground stars are now ignored by the astrometric routines. This may result in some small changes in astrometry for some objects.
\item The bias calculation in the blue camera has been improved. The impact of this will be negligible except to improve the quality of some extremely bright emission lines.
\item Adjustments were made to the sky subtraction algorithms to optimize performance for the Coma cluster ancillary program. 
\item There have been improvements in the data quality flagging for cubes with dead fibres. 
\end{itemize}

Instructions for accessing the MaNGA data products are given on the SDSS
website\footnote{\url{http://www.sdss.org/dr14/manga/manga-data/data-access/}}.  We summarize available options here and
refer the reader to the DR13 paper for additional details.  All data products are stored on the Science Archive Server
at \url{http://data.sdss.org/sas/dr14/manga/spectro/redux/}. Here you will find the {\tt drpall} summary table as well as subdirectories
that store reduction output for each plate, both for observations obtained on a specific night and for the results of
combining all observations of a given plate into a ``stack.''  The drpall table may be queried either after downloading
this file to disk or through the SDSS CASJobs system.  Such queries define selections of galaxies of interest and can
return the plate-IFU combination for those galaxies that identifies how they were observed.  These in turn can be used
to find the SAS directory locations of the corresponding data products.  Large downloads can be accomplished via rsync
calls as described on the SDSS website.  Finally, the SDSS SkyServer Explore tool provides basic information about
MaNGA targets.  

Several features of the MaNGA data should be kept in mind while using the data.  Most important, each MaNGA data cube
has a FITS header keyword DRP3QUAL indicating the quality of the reduction. 1-2\% of the data cubes are flagged as
significantly problematic---galaxies with CRITICAL quality bit (=30) set should be treated with extreme caution (see L16).  Please also use the MASK extension of each datacube to identify problematic spaxels.  A simple
summary {\tt DONOTUSE} bit is of particular importance indicating elements that should be masked out for scientific
analyses.  

There is significant covariance between adjacent spaxels in data cubes, given that the spaxel size (0.5$\arcsec$) is much smaller than the fiber size (2$\arcsec$ diameter).  A simple method that accounts for covariance when one desires to spatially bin spaxels together is discussed in \S 9.3 of L16. The typical reconstructed point
spread function of the MaNGA data cubes has a FWHM of 2.5$\arcsec$.  Sparse correlation matrices in the $ugriz$ central wavelengths are also now provided in the data cubes.

As discussed by L16, the instrumental line spread function (LSF) in the DR13 data was underestimated by about
$10\pm2$\%.  This has been corrected in DR14 and the reported LSF is described by a post-pixellized gaussian.

Additional issues and caveats are discussed in \url{http://www.sdss.org/dr14/manga/manga-caveats/}.

\subsection{Highlights of MaNGA Science with DR14 Data}

The MaNGA survey has produced a number of scientific results based on data acquired so far, indicating the breadth of research possible with the MaNGA data.  In the DR13 paper we provided a summary of science highlights with early data. Here we briefly summarize the results of papers that have been completed within the SDSS-IV collaboration using the MaNGA sample released as part of DR14. 

For example, published results based on the MaNGA DR14 data include, \citet{BBaccepted} who discuss the integrated stellar mass-metallicity relation for more than 1700 galaxies, \citet{Zhu17} who revisit the relation between the stellar surface density, the gas surface density and the gas-phase metallicity of typical disc galaxies in the local Universe, \citet{Belfiore17} who study the gas phase metallicity and nitrogen abundance gradients traced by star-forming regions in a representative sample of 550 nearby galaxies, and \citet{2017ApJ...837...32L} who report the discovery of a mysterious giant H$\alpha$ blob that is $\sim 8$ kpc away from a component of a dry galaxy merger. In \citet{Bizyaev17} was presented a study of the kinematics of the extraplanar ionized gas around several dozen galaxies, while \citet{2017A&A...599A.141J} conducted a detailed study of extra-planar diffuse ionized gas stacking spectra from 49 edge-on, late-type galaxies as a function of distance from the midplane of the galaxy. Numerous other results based on DR14 data are in preparation. 

\subsection{MaNGA VACs}

This data release also contains two VACs based on MaNGA data. They are briefly summarized below, and for more details we refer you to the papers given in Table \ref{table:vac}

\subsubsection{MaNGA Pipe3D value added catalog: Spatially resolved and integrated properties of galaxies}
\label{sec:mangavac1}

{\sc Pipe3D} is an IFU-focused analysis pipeline that calculates intermediate dataproducts and is able to obtain both the stellar population and the ionized gas properties extracted from the datacubes in an automatic way. This pipeline is based on {\sc FIT3D}, details of which are presented in \citet{2016RMxAA..52...21S,2016RMxAA..52..171S} which show some examples based on CALIFA \citep{2016A&A...587A..70S,2016ApJ...821L..26C,2017MNRAS.469.2121S} and MaNGA/P-MaNGA \citep{2016MNRAS.463.2799I,2016MNRAS.463.2513B,2017ApJ...837...32L,BBaccepted} datasets. The MaNGA dataproducts provided by Pipe3D are presented in \citet{2017arXiv170905438S}\footnote{\url{http://www.sdss.org/dr14/manga/manga-data/manga-pipe3d-value-added-catalog}}. The VAC consists of a single table containing integrated (cumulative), characteristic (values at the effective radius), and gradient of different quantities, included stellar mass, star-formation (and their densities), oxygen and nitrogen abundances, dust attenuation, estimated gas density, stellar and gas velocity dispersions.

For each galaxy, data is presented as an individual FITS file including four extensions, each one corresponding to a data cube that comprises (1) the spatial resolved properties required to recover the star-formation histories, (2) the average properties of the stellar populations, (3) the emission line properties for
56 strong and weak emission lines (including the former ones together with the EW of the lines), and (4)  the most frequently used stellar indices. The details of each individual extension was described in S\'anchez et al. (2016b), and the final adopted format in  S\'anchez et al. (in prep.).

\subsubsection{MaNGA FIREFLY Stellar Populations}
\label{sec:mangavac2}

The MaNGA {\sc FIREFLY} VAC \citep{2017MNRAS.466.4731G} provides measurements of spatially resolved stellar population properties in MaNGA galaxies. It is built on and complements the products of the MaNGA data analysis pipeline (DAP, Westfall et al, in preparation) by providing higher-order and model-based data products. These are measurements of optical absorption line-strengths, as well as the physical properties age, metallicity and dust attenuation. The latter are derived from full spectral fitting with the code FIREFLY \citep{2015MNRAS.449..328W,2017MNRAS.472.4297W} using the supercomputer {\sc SCIAMA2} at Portsmouth University. The VAC is a single fits file (4 GB) containing measurements all DR14 MaNGA galaxies. The catalogue contains basic galaxy information from the literature (i.e., galaxy identifiers, redshift, mass), global derived parameters (i.e., light-weighted and mass-weighted stellar population ages and metallicities for a central 3 arcsec aperture and for an elliptical shell at 1 effective radius), gradient parameters (i.e., gradients in age and metallicity measured within $1.5\; R_{\rm e}$) and spatially resolved quantities (i.e., 2-D maps of age, metallicity, dust attenuation, mass and surface mass density, and 28 absorption line indices).

More detail on the catalogue, and the method of creating the 2-dimensional maps is provided in \citet{2017MNRAS.466.4731G} and the data is available from the data release website\footnote{\url{http://www.sdss.org/dr14/manga/manga-data/manga-firefly-value-added-catalog}}.

\section{Future Plans}
\label{sec:future}

SDSS-IV is planning a 6-year survey, with operations at both the 2.5 meter Sloan Foundation Telescope at Apache Point Observatory, New Mexico, USA and the du Pont Telescope at Las Campanas, Chile scheduled through 2020. Future data releases from SDSS-IV will include data observed with both telescopes;  the final SDSS-IV data release is planned to be DR18, currently scheduled for December 2020. 

For APOGEE, future data releases will include, for the first time, southern hemisphere observations taken with the new APOGEE-S instrument at the Las Campanas Observatory with the duPont 2.5m telescope. These observations will extend APOGEE coverage to the full Galaxy, with significantly increased observations of the Galactic bulge and also include observations in the Magellanic Clouds, globular clusters, and dwarf spheroidal galaxies only accessible from the southern hemisphere. As usual, future data releases will also include re-reductions of all APOGEE-N data. Plans for improved stellar parameter/abundance analysis include using a new homogeneous grid of MARCS stellar atmospheres, and the use of ``minigrids" to analyze elements whose absorption features are too blended with those of other elements to be reliably extracted with the abundance techniques used to date. 

For MaNGA, is is planned that the DR15 data release will include the $\sim$4000 MaNGA galaxies which have been observed up to summer shutdown 2017.  In addition, we anticipate a number of new data products to be released in this and future DRs.  These include reduced spectra from the MaStar stellar library (Yan et al. in prep.), which is making use of commensal observations during APOGEE-2 time to obtain spectroscopic observations of stars which will be used to build a new stellar library through the MaNGA instrumentation; and output from the MaNGA Data Analysis Pipeline (DAP; Westfall et al. in prep.).  The DAP produces maps maps of emission line fluxes, gas and stellar kinematics, and stellar population properties.  Some similar derived data products are already available as
Value Added Catalogs (see Table \ref{table:vac} and Section \ref{sec:mangavac1} and \ref{sec:mangavac2}).  Finally, we intend for DR15 to mark the first release of the ``Marvin'' ecosystem which includes powerful python tools for seamlessly downloading and querying the MaNGA data as well as web interface
that provides advanced search functionality, an user interface to the MaNGA datacubes, and the ability to quickly
choose and display maps of key quantities measured by the DAP.

For eBOSS, future data releases will include the ELG survey results as well as the continuation of the LRG-QSO surveys. They will also include further value added catalogues: in particular the continuation of the quasar catalogue, a detailed ELG catalogue, as well as large scale structure clustering catalogues required for independent clustering analysis. Further improvement on the redshift measurement and spectral classification catalogue is also likely.

For TDSS, a future SDSS data release will include very recent spectra from its Repeat Quasar Spectroscopy (RQS) program, which obtains multi-epoch
spectra for thousands of known quasars, all of which have least one epoch of SDSS spectroscopy available (and often already archived). Quasar spectral variability on multi-year timescales is currently poorly characterized for large samples athough there are many exciting results from smaller select subsets (see \citealt{Runnoe_etal_2016} and \citealt{McGraw_etal_2017} for examples of studies based on repeat spectroscopy, ranging from discoveries of new changing look quasars, to broad absorption line or BAL emergence and disappearance). The RQS program in TDSS will ultimately observe $\sim10^4$ known (SDSS) quasars in the ELG survey region \citep{2017arXiv170400338R} adding at least one additional spectral
epoch. This will allow for an extension of earlier work to a systematic investigation of quasar spectroscopic variability, both by making a larger sample, but also by including large numbers of quasars as targets for repeat spectra that were selected without {\it a priori} knowledge of their specific quasar spectroscopic sub-class or variability properties. A recent detailed technical description of target selection for all of the TDSS repeat spectroscopy programs (including RQS), may be found in \citet{MacLeod_etal_2017}. 

For SPIDERS, future data releases will focus on higher-level data products, such as black hole masses and host galaxy properties of the X-ray AGN, as well as rich characterizarion of the X-ray selected clusters (in particular, dynamical properties and calibrated cluster masses). The first spectra of counterparts of eROSITA sources, however, will only be obtained beginning in Spring 2019, so they will be part of DR18 and subsequent releases only.

Planning has begun for the next generation of SDSS, to begin in 2020 \citep{2017arXiv171103234K}. SDSS-V will build on the SDSS infrastructure and expand the instrumentation (especially for optical IFU spectroscopy) in both hemispheres. This expansion of SDSS's legacy will enable an enormous sample comprising millions of spectra of quasars, galaxies, and stars, with scientific goals ranging from the growth of supermassive black holes to the chemical and dynamical structure of the Milky Way, the detailed architecture of planetary systems and the astrophysics of star formation.

\section{Acknowledgements}

We would like to thank the University of St Andrews, Scotland for their hospitality during DocuCeilidh 2017.

Funding for the Sloan Digital Sky Survey IV has been provided by
the Alfred P. Sloan Foundation, the U.S. Department of Energy Office of
Science, and the Participating Institutions. SDSS-IV acknowledges
support and resources from the Center for High-Performance Computing at
the University of Utah. The SDSS web site is www.sdss.org.

SDSS-IV is managed by the Astrophysical Research Consortium for the 
Participating Institutions of the SDSS Collaboration including the 
Brazilian Participation Group, the Carnegie Institution for Science, 
Carnegie Mellon University, the Chilean Participation Group, the French Participation Group, Harvard-Smithsonian Center for Astrophysics, 
Instituto de Astrof\'isica de Canarias, The Johns Hopkins University, 
Kavli Institute for the Physics and Mathematics of the Universe (IPMU) / 
University of Tokyo, Lawrence Berkeley National Laboratory, 
Leibniz Institut f\"ur Astrophysik Potsdam (AIP),  
Max-Planck-Institut f\"ur Astronomie (MPIA Heidelberg), 
Max-Planck-Institut f\"ur Astrophysik (MPA Garching), 
Max-Planck-Institut f\"ur Extraterrestrische Physik (MPE), 
National Astronomical Observatories of China, New Mexico State University, 
New York University, University of Notre Dame, 
Observat\'ario Nacional / MCTI, The Ohio State University, 
Pennsylvania State University, Shanghai Astronomical Observatory, 
United Kingdom Participation Group,
Universidad Nacional Aut\'onoma de M\'exico, University of Arizona, 
University of Colorado Boulder, University of Oxford, University of Portsmouth, 
University of Utah, University of Virginia, University of Washington, University of Wisconsin, 
Vanderbilt University, and Yale University.

\bibliographystyle{apj}

\end{document}